\documentclass[runningheads]{llncs}

\usepackage[T1]{fontenc}
\usepackage{graphicx}
\usepackage{subcaption}
\usepackage[colorlinks=true, allcolors=blue]{hyperref}
\usepackage{CJK}
\usepackage{amsmath}
\usepackage{multirow}
\usepackage{diagbox}
\usepackage{marvosym}
\usepackage{booktabs}

\title{The First Voice Timbre Attribute Detection Challenge \thanks{This work was supported in part by the National Key Research and Development Program of China Project 2024YFE0217200, the Innovation and Technology Fund of the Hong Kong SAR MHP/048/24, the National Natural Science Foundation of China under Grant U23B2053, and the Fundamental Research Funds for the Central Universities WK2100000043.}}

\author{Liping Chen\inst{1}
\and
Jinghao He\inst{1} \and
Zhengyan Sheng\inst{1}\and \\ Kong Aik Lee \inst{2} \and Zhen-Hua Ling \inst{1}}
\authorrunning{L. Chen et al.}
\institute{University of Science and Technology of China, Hefei, China \\ \email{\{lipchen, zhling\}@ustc.edu.cn, \{zysheng, jhhe\}@mail.ustc.edu.cn} \and
The Hong Kong Polytechnic University, Hong Kong, China \email{kong-aik.lee@polyu.edu.hk}}

\begin{document}
\maketitle              % typeset the header of the contribution
\begin{abstract}
The first voice timbre attribute detection challenge is featured in a special session at NCMMSC 2025. It focuses on the explainability of voice timbre and compares the intensity of two speech utterances in a specified timbre descriptor dimension. The evaluation was conducted on the VCTK-RVA dataset. Participants developed their systems and submitted their outputs to the organizer, who evaluated the performance and sent feedback to them. Six teams submitted their outputs, with five providing descriptions of their methodologies.

\end{abstract}

\section{Introduction}
Voice timbre, or voice quality, is inherently perceptual, reflecting the psychological impression generated by a physical stimulus, and is determined by both the listener and the voice \cite{handbook}. It has been widely researched across disciplines such as acoustics \cite{acoustic_parameter}, medicine \cite{clinical_practice}, and psychology \cite{incidental}. Specifically, within acoustics, voice recognition has been extensively studied, primarily focusing on the similarity between voice characteristics \cite{perceptual_dissimilarity}. Recently, voice explainability has emerged as a critical factor in speech technologies, such as speech generation \cite{vctk-rva} and forensic voice analysis \cite{wu2024explainable}.

The first voice timbre attribute detection (vTAD) challenge \cite{sheng2025voicetimbreattributedetection} was held at the National Conference on Man-Machine Speech Communication 2025 (NCMMSC2025) conference, with the objective of explaining voice timbre from the perspective of human impression. In this challenge, a set of timbre descriptors derived from sensory attributes across various modalities was built, including sound (hoarse, rich), vision (bright, dark), texture (soft, hard), physical attribute (magnetic, transparent), and so on. Given a pair of speech utterances and a specified descriptor dimension, systems were developed to compare the intensity of the two utterances along that dimension. Finally, in this challenge, six teams submitted their outputs, with five providing descriptions of their methodologies. This paper summarizes the NCMMSC2025-vTAD challenge, including the task and submitted systems.

\section{Task}
This section describes the challenge task in terms of task definition, dataset, and evaluation metrics.

\subsection{Task definition}
In this task, a set of timbre descriptors \(\mathcal{V}\) was defined. Given two utterances from distinct speakers A and B, denoted as \({\mathcal O}_{\rm A}\) and \({\mathcal O}_{\rm B}\), and a specified timbre descriptor v \( \in \mathcal{V}\), participants were required to develop one or more vTAD systems to compare the intensity of \({\mathcal O}_{\rm A}\) and \({\mathcal O}_{\rm B}\) in v. 

Mathematically, a hypothesis about the intensity difference was defined as ${\mathcal H}\left(\langle{\mathcal O}_{\rm A}, {\mathcal O}_{\rm B}\rangle, {\rm v}\right)$, meaning that ${\mathcal O}_{\rm B}$ was stronger than ${\mathcal O}_{\rm A}$ in the descriptor dimension v. Specifically, ${\mathcal H} \in \{0, 1\}$, where ${\mathcal H} = 1$ indicated that the hypothesis ${\mathcal H}$ was correct, and ${\mathcal H} = 0$ indicated that the hypothesis was incorrect. Participants were required to provide two types of output: the score \(s_{\langle{\rm A},{\rm B}\rangle}^{\rm v}\), indicating the likelihood that the hypothesis \({\mathcal{H}\left(\langle{\mathcal O}_{\rm A},{\mathcal O}_{\rm B}\rangle, {\rm v}\right)}=1\) was true, and a decision about whether the hypothesis \({\mathcal{H}\left(\langle{\mathcal O}_{\rm A},{\mathcal O}_{\rm B}\rangle, {\rm v}\right)}=1\) was true. The likelihood scores and decisions were measured with equal error rate (EER) and accuracy (ACC), respectively.

Moreover, in this challenge, two evaluation tracks were defined, including \emph{unseen} and \emph{seen}. In the unseen scenario, the speakers used in the evaluation phase were not present in the training phase. In the seen scenario, the speakers employed in the evaluation phase were applied in the training phase, while distinct utterances were utilized for training and evaluation, respectively. Moreover, given a specific speaker, the ordered pairs composed with different speakers were exclusively used in training and evaluation.

\subsection{Dataset}
The challenge utilized the VCTK-RVA dataset \cite{vctk-rva}, wherein the publicly available VCTK database was annotated for timbre intensity. In the dataset, 18 timbre descriptors were defined in $\mathcal V$, as listed in Table \ref{tab: vctk-rva dataset}. In total, 101 speakers were involved, forming 6,038 annotated ordered speaker pairs \{Speaker A, Speaker B, voice attribute v\}, indicating that Speaker B was stronger than Speaker A in the specific descriptor v. The number of descriptor dimensions annotated for each ordered speaker pair ranged from 1 to 3.

\begin{table}[t]
	\centering
		\caption{The descriptor set used for describing the timbre. The \emph{Trans.} column gives the corresponding Chinese word. The \emph{Perc.} column presents the percentage (\%) of each descriptor in the \emph{VCTK-RVA} dataset. The descriptors \emph{shrill} and \emph{husky} are exclusively annotated for female and male, respectively.}
	\begin{tabular}{lcc lcc}
		\toprule[1pt]
		
		\textbf{Descriptor} & \textbf{Trans.} & \textbf{Perc.} & \textbf{Descriptor} & \textbf{Trans.} & \textbf{Perc.} \\ 
		\cmidrule(lr){1-3} \cmidrule(lr){4-6}
		Bright      & \begin{CJK*}{UTF8}{gbsn}明亮\end{CJK*}      & 17.10      & Thin  &\begin{CJK*}{UTF8}{gbsn}单薄\end{CJK*}     & 13.03      \\
		Coarse &\begin{CJK*}{UTF8}{gbsn}粗\end{CJK*}      & 11.62      & Slim  &\begin{CJK*}{UTF8}{gbsn}细\end{CJK*}      & 11.31      \\
		Low &\begin{CJK*}{UTF8}{gbsn}低沉\end{CJK*}        & 7.43       & Pure &\begin{CJK*}{UTF8}{gbsn}干净\end{CJK*}       & 5.48       \\
		Rich &\begin{CJK*}{UTF8}{gbsn}厚实\end{CJK*}       & 4.71       & Magnetic  &\begin{CJK*}{UTF8}{gbsn}磁性\end{CJK*}  & 3.64       \\
		Muddy  &\begin{CJK*}{UTF8}{gbsn}浑浊\end{CJK*}     & 3.59       & Hoarse &\begin{CJK*}{UTF8}{gbsn}沙哑\end{CJK*}     & 3.32       \\
		Round  &\begin{CJK*}{UTF8}{gbsn}圆润\end{CJK*}     & 2.48       & Flat  &\begin{CJK*}{UTF8}{gbsn}平淡\end{CJK*}      & 2.15       \\
		Shrill  (female only) &\begin{CJK*}{UTF8}{gbsn}尖锐\end{CJK*}     & 2.08       & Shriveled &\begin{CJK*}{UTF8}{gbsn}干瘪\end{CJK*}  & 1.74       \\
		Muffled &\begin{CJK*}{UTF8}{gbsn}沉闷\end{CJK*}    & 1.44       & Soft    &\begin{CJK*}{UTF8}{gbsn}柔和\end{CJK*}    & 0.82       \\
		Transparent &\begin{CJK*}{UTF8}{gbsn}通透\end{CJK*} & 0.66       & Husky (male only)&\begin{CJK*}{UTF8}{gbsn}干哑\end{CJK*}      & 0.59       \\ 
		\bottomrule[1pt]
	\end{tabular}

	\label{tab: vctk-rva dataset}
\end{table}

In this challenge, the VCTK-RVA dataset was partitioned for training and evaluation, respectively. The evaluations were defined in two tracks regarding whether the test speakers were seen or not in the training. In the seen evaluation track, the speakers were included in the training data, while the evaluated speaker pairs were not. For each gender, the training set contained speaker pairs annotated on all 17 voice attributes. In total, 29 male and 49 female speakers were included in the training phase. In the testing phase, five descriptors were selected for each gender. In detail, male utterances were examined on descriptors \emph{Bright}, \emph{Thin}, \emph{Low}, \emph{Magnetic}, and \emph{Pure}, while female utterances were examined on descriptors \emph{Bright}, \emph{Thin}, \emph{Low}, \emph{Coarse}, and \emph{Slim}. In both evaluation datasets, 20 utterances were randomly drawn from each speaker. For a speaker pair, 100 ${\mathcal H}=1$ and 300 ${\mathcal H}=0$ evaluation samples were constructed for a descriptor.

\subsection{Baseline method}
\begin{figure}[t]
\centering
\begin{subfigure}[t]{1\textwidth}
        \centering
        \includegraphics[scale=1.0]{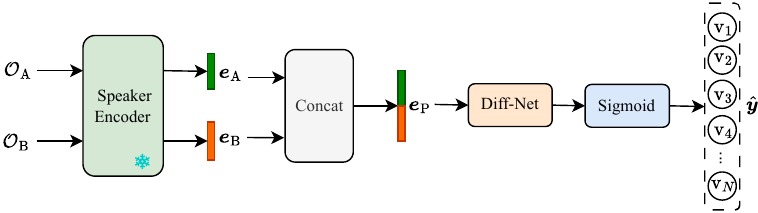}
        \caption{Overall architecture}
        \label{fig: overall architecture}
    \end{subfigure}
    \hfill
    \vspace{0.4cm}
    \begin{subfigure}[t]{1\textwidth}
        \centering
        \includegraphics[scale=1.0]{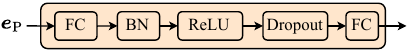}
        \caption{Diff-Net}
        \label{fig: diff-net}
    \end{subfigure}
    \caption{Baseline method in NCMMSC2025-vTAD challenge. The speaker encoder was pre-trained and frozen. The notation Conca is short for concatenation, FC is short for fully-connected layer, BN is short for batch normalization.}
\label{fig: model architecture}
\end{figure}

The challenge utilized the methodology outlined in \cite{baselinepaper} as the baseline. As depicted in Fig. \ref{fig: model architecture}, in the baseline method, speaker embedding vectors were extracted from the utterance pair ${\mathcal O}_{\rm A}$ and ${\mathcal O}_{\rm B}$ with a speaker encoder, and represented with ${\boldsymbol{e}}_{\rm A}$ and ${\boldsymbol{e}}_{\rm B}$, respectively. The concatenation of both was obtained as $e_{\rm P}$, which was then input to the Diff-Net module. The output dimension of the Diff-Net was $N$, equal to the number of descriptors defined in ${\mathcal{V}}$. Thereafter, the sigmoid function was applied to each node, producing the vector \({\hat{\boldsymbol{y}}}\), where the \(n\)-th $\left(n=1,2,...,N\right)$ dimension was the prediction of the intensity comparison for the corresponding descriptor. In this challenge, the speaker encoder utilized pre-trained models, specified as ECAPA-TDNN \cite{desplanques2020ecapa} and FACodec \cite{facodec}, which were frozen during model training.

\section{Submitted systems}
\begin{figure}[t]
\centering
\begin{subfigure}[t]{1\textwidth}
        \centering
        \includegraphics[scale=0.5]{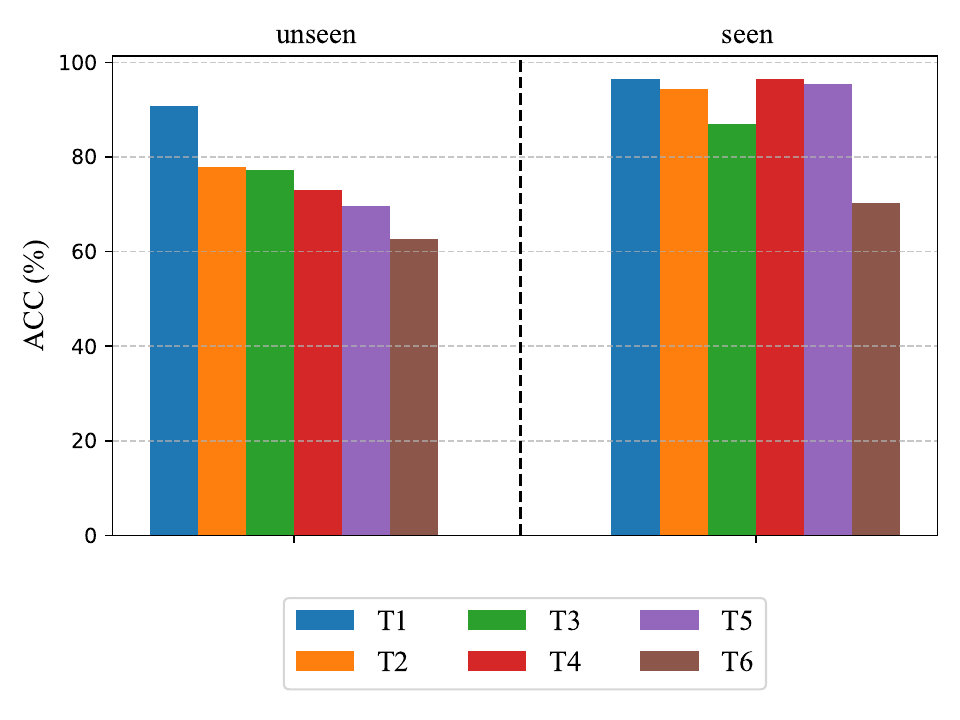}
        \caption{ACC}
        \label{fig: ACC res}
    \end{subfigure}
    \hfill
    \vspace{0.4cm}
    \begin{subfigure}[t]{1\textwidth}
        \centering
        \includegraphics[scale=0.5]{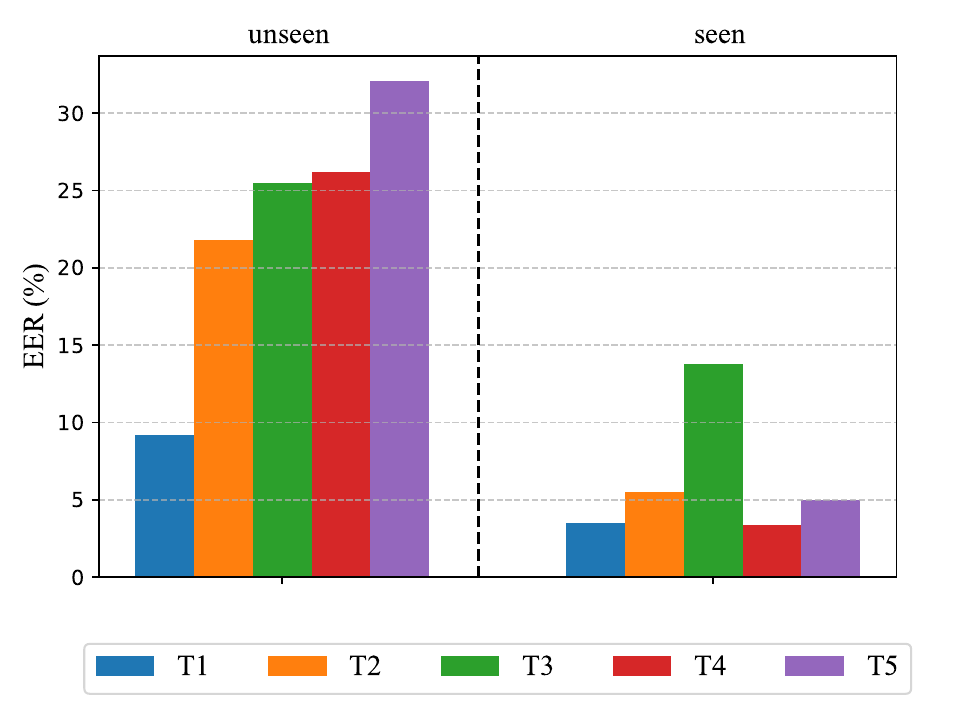}
        \caption{EER}
        \label{fig: EER res}
    \end{subfigure}
    \caption{Accuracy (ACC) obtained by six teams and EER obtained by five teams in unseen and seen evaluation tracks, respectively.}
\label{fig: res}
\end{figure}

The challenge attracted 32 registered teams from academia and industry in 5 countries. Among them, 6 teams successfully submitted their results, represented as T1 to T6. Among them, five teams submitted both likelihood scores and decisions, while one team submitted only decisions. For each team, the system exhibiting the best performance was utilized for ranking. Specifically, the EER and ACC were respectively obtained as the average across all the descriptors from both genders, which are summarized in Fig. \ref{fig: res}.

From the results, it can be seen that T1 achieved the best performance in the unseen track, indicated by the highest accuracy and lowest EER. In the seen track, T4 provided the best system, as evidenced by its highest accuracy and lowest EER among all the submitted systems.

Particularly in terms of methodologies, T2 and T4 examined the speaker encoder module for the extraction of timbre representation, wherein the SiamAM-ResNet \cite{SiamAM-ResNet} and WavLM \cite{WavLM} were explored. T1, T2, T3, and T5 investigated the implementation of the Diff-Net function, based on the embedding vectors derived from the input utterance pair. Moreover, T3 considered the data imbalance problem of the descriptors and integrated a graph-based model for data augmentation. Readers are suggested to refer to the special session for NCMMSC2025-vTAD challenge in the proceedings of NCMMSC2025 for further details.

\bibliographystyle{unsrt}
\bibliography{myref}

\begin{thebibliography}{10}

\bibitem{handbook}
D.~B. Pisoni and R.~E. Remez.
\newblock {\em The Handbook of Speech Perception}.
\newblock Blackwell Publishing Ltd, USA, 2005.

\bibitem{acoustic_parameter}
W.~A. van Dommelen.
\newblock Acoustic parameters in human speaker recognition.
\newblock {\em Language and Speech}, 33(3):259--272, 1990.

\bibitem{clinical_practice}
F.~Robert et~al.
\newblock The perceived role of voice perception in clinical practice.
\newblock {\em PHONOSCOPE}, 2(2):87--106, 1999.

\bibitem{incidental}
R.~E. Geiselman and F.~S. Bellezza.
\newblock Incidental retention of speaker\'s voice.
\newblock {\em Memory \& Cognition}, 5:658--65, 1977.

\bibitem{perceptual_dissimilarity}
T.~K. Perrachione, K.~T. Furbeck, and E.~J. Thurston.
\newblock Acoustic and linguistic factors affecting perceptual dissimilarity judgments of voices.
\newblock {\em The Journal of the Acoustic Society of America}, 146(5):3384--3399, 2019.

\bibitem{vctk-rva}
Z.-Y. Sheng, L.-J. Liu, Y.~Ai, J.~Pan, and Z.-H. Ling.
\newblock Voice attribute editing with text prompt.
\newblock {\em IEEE Transactions on Audio, Speech and Language Processing}, 33:1641--1652, 2025.

\bibitem{wu2024explainable}
X.~Wu, C.~Luu, P.~Bell, and R.~Ajitha.
\newblock Explainable attribute-based speaker verification.
\newblock {\em CoRR}, abs/2405.19796, 2024.

\bibitem{sheng2025voicetimbreattributedetection}
Z.~Sheng, J.~He, L.~Chen, K.~A. Lee, and Z.-H. Ling.
\newblock The voice timbre attribute detection 2025 challenge evaluation plan.
\newblock {\em arXiv}, 2025.

\bibitem{baselinepaper}
J.~He, Z.~Sheng, L.~Chen, K.~A. Lee, and Z.-H. Ling.
\newblock Introducing voice timbre attribute detection.

\bibitem{desplanques2020ecapa}
B.~Desplanques, J.~Thienpondt, and K.~Demuynck.
\newblock {ECAPA-TDNN}: Emphasized channel attention, propagation and aggregation in {TDNN} based speaker verification.
\newblock In {\em Proc. Interspeech}, pages 3830--3834, 2020.

\bibitem{facodec}
Z.~Ju, Y.~Wang, K.~Shen, et~al.
\newblock Naturalspeech 3: Zero-shot speech synthesis with factorized codec and diffusion models.
\newblock In {\em Proc. {ICML}}, 2024.

\bibitem{SiamAM-ResNet}
X.~Qin, N.~Li, C.~Weng, D.~Su, and M.~Li.
\newblock Simple attention module based speaker verification with iterative noisy label detection.
\newblock In {\em Proc. ICASSP}, pages 6722--6726, 2022.

\bibitem{WavLM}
S.~Chen et~al.
\newblock {WavLM}: Large-scale self-supervised pre-training for full stack speech processing.
\newblock 16(6):1505--1518, 2022.

\end{thebibliography}

\end{document}